\documentclass[conference]{IEEEtran}

\usepackage[pdftex]{graphicx}
\usepackage{amsmath}
\usepackage{eqparbox}
\usepackage{algorithmic}
\usepackage{xcolor}
\usepackage{comment}
\usepackage[font=small,skip=0pt]{caption}
\usepackage{cite}
\usepackage{adjustbox}
\usepackage{hyperref}
\usepackage[affil-it]{authblk}
\begin{document}

%+++++++++++++++++++++++++++++++++++++++++++++++++++
\title{\LARGE Accelerating CNN inference on long vector architectures via co-design}
%Interplay of CNN algorithmic optimizations and long vector architectures
%Co-design exploration of vector ISAs and microarchitectures for optimized CNNs with long vectors
\author{Sonia Rani Gupta, Nikela Papadopoulou, Miquel Pericas\\
{\{soniar, nikela, miquelp\}}@chalmers.se}
\affil{Dept Computer Science and Engineering, \\Chalmers University Of Technology, Sweden}

%\title{\LARGE HW/SW codesign of Vector Architectures and CNNs}
%\author{\authorblockN{Leave Author List blank for your IMS2014 Summary (initial) submission.\\ IMS2014 will be rigorously enforcing the double-blind reviewing requirements.}
%\authorblockA{\authorrefmark{1}Leave Affiliation List blank for your Summary (initial) submission}}
%+++++++++++++++++++++++++++++++++++++++++++++++++++

\maketitle

%\thispagestyle{plain}
%\pagestyle{plain}
% ================
% # Abstract     #
% ================

\begin{abstract}
%CPU-based inference can be deployed as an alternative to off-chip accelerators. In this context, emerging vector architectures are a promising option, owing to their high efficiency. Yet the large design space of convolution algorithms and hardware implementations makes the selection of design options challenging. In this paper we present our ongoing research into co-designing future vector architectures for CPU-based convolutional Neural Networks (CNN) inference focusing on the im2col+GEMM and Winograd kernels. Using the Gem5 simulator we explore the impact of several hardware microarchitectural features including (i) vector lanes, (ii) vector lengths, (iii) cache sizes, and (iv) options for integrating the vector unit into the CPU pipeline. In the context of im2col+GEMM, we consider several BLIS-like algorithmic optimizations such as (1) utilization of vector registers, (2) loop unrolling and (3) loop reorder, (4) manual vectorization, (5) prefetching, and (6) packing of matrices, on RISC-V Vector and ARM-SVE. Finally, we describe our approach of inter-tile parallelization across input/output channels for the vectorization of the Winograd convolution on ARM-SVE. Our method provides good memory reuse and exploitation of longer vector lengths.
CPU-based inference can be deployed as an alternative to off-chip accelerators. In this context, emerging vector architectures are a promising option, owing to their high efficiency. Yet the large design space of convolutional algorithms and hardware implementations makes the selection of design options challenging. In this paper, we present our ongoing research into co-designing future vector architectures for CPU-based Convolutional Neural Networks (CNN) inference focusing on the im2col+GEMM and Winograd kernels. Using the Gem5 simulator we explore the impact of several hardware microarchitectural features including (i) vector lanes, (ii) vector lengths, (iii) cache sizes, and (iv) options for integrating the vector unit into the CPU pipeline. 
In the context of im2col+GEMM, we study the impact of several BLIS-like algorithmic optimizations such as (1) utilization of vector registers, (2) loop unrolling, (3) loop reorder, (4) manual vectorization, (5) prefetching, and (6) packing of matrices, on RISC-V Vector and ARM-SVE ISAs. We use the YOLOv3 and VGG16 network models for our evaluation. Our co-design study shows that BLIS-like optimizations are not beneficial to all types of vector microarchitectures.  
We additionally demonstrate that longer vector lengths (of at least 8192 bits) and larger caches (of 256MB) can boost performance by 5$\times$, with our optimized CNN kernels, compared to a vector length of 512-bit and 1MB of L2 cache. 
In the context of Winograd, we present our novel approach of inter-tile parallelization across the input/output channels by using 8$\times$8 tiles per channel to vectorize the algorithm on vector length agnostic (VLA) architectures. Our method exploits longer vector lengths and offers high memory reuse, resulting in performance improvement of up to 2.4$\times$ for non-strided convolutional layers with 3$\times$3 kernel size, compared to our optimized im2col+GEMM approach on the Fujitsu A64FX processor. Our co-design study furthermore reveals that Winograd requires smaller cache sizes (up to 64MB) compared to im2col+GEMM.

%Our study provides the performance evaluation after applying these optimizations on a Fujitsu A64FX processor. 
\end{abstract}

\section{Introduction}
\label{sec:introduction}
% ========================
% # I. Introduction      #
% ========================

Inference via Convolutional Neural Networks (CNNs) is used in many AI applications such as object detection~\cite{objectdetection}, natural language processing~\cite{NLP} and speech recognition~\cite{speech}. Most CNN-based object detection network models work with a tight response-time limit and have high and increasing computation costs~\cite{objectdetectionnetworkmodels,computeintensive,yolov3}.
%\remove{ The computational cost of these CNN based network models is dominated by the computation required to perform the convolutions over tensors~\cite{computeintensive,yolov3}, and grows over time.}
%For example, Resnet152 and YoloV3 work with 11 billion and 60 billion floating-point operations respectively.
Additionally, these models often operate under tight power constraints, e.g. battery power in embedded systems~\cite{powerconstrainedonmobile}, or power caps in datacenters~\cite{ powercapdatacenter}. Therefore, highly accurate real-time CNNs require highly optimized kernels, running on energy-efficient architectures with large computational capacity. 

The popular approach for CNN inference, adopted by many frameworks~\cite{tensorflow, PyTorch, yolov3} is to offload the compute-intensive kernels to GPUs~\cite{neuralacce, gpucnn,intgrategpu}. Specialized neural accelerators also exist \cite{abdelouahab2018accelerating, moolchandani2021accelerating}, but their integration in the general-purpose computing stack is challenging. 
%Offloading the compute-intensive parts to dedicated neural accelerators and general-purpose GPUs has emerged as a solution for accelerating CNN inference~\cite{neuralacce}. While specialized neural accelerators, such as FPGAs \\cite{abdelouahab2018accelerating} and ASICs \cite{moolchandani2021accelerating}, are effective for deep learning, their integration into the general-purpose computing stack is challenging. 
%{\remove {Various Deep Neural Network (DNN) frameworks such as TensorFlow~\cite{tensorflow}, PyTorch \cite{Pytorch}, or Darknet \cite{yolov3}, are optimized for computing platforms coupled with discrete GPUs}}.
%Caffe \cite{caffe}, 
Nevertheless, many use cases require availability, low-latency, or portability \cite{FBdatacenter, mittaldnnCPU, optimizingcnnoncpu}, and therefore benefit from executing deep neural networks (DNNs) on tightly integrated systems. 
%Although CNN accelerators exist, several use cases requiring availability, low-latency, and portability \cite{FBdatacenter, mittaldnnCPU, optimizingcnnoncpu} benefit from executing deep neural networks (DNNs) on tightly integrated systems. 
Consequently, many works target software optimization of CNN inference on CPUs \cite{optimizingcnnoncpu,georganasTensorDNN, RuiliAnalyticalOpt}, while CPU vendors increasingly add DNN capabilities to processors~\cite{nextplatform1}. 
%\sonia{ On Server-side CNN inference workloads need availability and low latency~\cite{FBdatacenter} and in the context of mobile/embedded computing, CPUs can offer the required high availability and portability~\cite{mittaldnnCPU,optimizingcnnoncpu}. }
%\remove{In fact, multiple works target software optimization of CNN inference on CPUs~\cite{optimizingcnnoncpu,georganasTensorDNN, RuiliAnalyticalOpt}.} 
%However, the merits of using CPUs for CNN inference are further highlighted by the increasing addition of vector and/or matrix units and instruction sets for the acceleration of machine learning workloads by CPU vendors.

In this aspect, %the recently re-introduced 
vector processors play a leading role~\cite{vectoracceEnegy}. Contemporary vector architectures, such as ARM-SVE~\cite{ARMSVE} and the RISC-V Vector extension~\cite{EPILink} 
%\remove{, adopted by several projects and products (e.g., EPI~\cite{EPILink}, Fujitsu A64FX~\cite{Fugaku}, ARM Neoverse-V1\cite{noverse-1}),} 
specify a maximum length of vector registers and allow the usage of different vector lengths. These vector-length agnostic (VLA) Instruction Set Architecture (ISAs) facilitate code portability across iterations of the same machine with different vector lengths. 
%\remove{ These vector processors can offer high performance to machine-learning workloads, and higher energy efficiency, making CPUs suitable for CNN inference.} 
%Vector processors based on VLA are featured prominently in various projects and products such as EPI~\cite{EPILink}, Fujitsu A64FX~\cite{Fugaku} and ARM Neoverse-V1\cite{noverse-1}.

%Kernel transformations are needed to extract the parallelism from the kernels to exploit the vector-unit power. Therefore, the effectiveness of vector processors depends not only on the hardware design, but also on algorithmic transformations and compiler optimizations.  Highly optimized kernel transformations are required for achieving high performance from the vector architectures. 
%porpodas2018autovectorization - autovectorization reference for space saving
The effectiveness of vector processors depends on algorithmic optimizations and the hardware design. First, given the limited compiler's ability to perform transformations during auto-vectorization \cite{ Adit2022microautovect}, transformations and optimizations to expose the available SIMD parallelism to the vector processing units are key to achieving high performance on vector architectures. Second, modern architectures can combine very long vector units, more on-chip vector parallelism and large caches. Tuning the micro-architectural parameters to the requirements of the vectorized and optimized kernels is integral to the design of high-performing, efficient vector architectures. 

%such as vector lenghts, vector lanes and cache sizes on the vector architectures is needed. Modern architectures can work with very long vector lengths, can have more on-chip vector parallelism and large cache sizes. Therefore, it is important to tune these hardware parameters in order to design a efficient vector architecture

Existing work on CNN inference on vector architectures focuses either on applying algorithmic optimizations \cite{ tensorlite,cococciono2020dnnsve,ARMComputeLibrary,gemmarm-sve} or on tuning the hardware micro-architectural parameters \cite{gem5benchmark, ARMSVEAnalysisVL,processcachestudy}. We identify the absence of a combined study as a missed opportunity to uncover algorithmic and architectural trade-offs in the performance of CNN kernels running on vector architectures.  In this work, we bridge this gap with a co-design study that performs a joint exploration of the design space of vector architectures and the optimization space of CNNs, aiming to provide guidance to programmers, hardware designers, and compiler developers. 

%These studies miss an opportunity to study and uncover the trade-offs in algorithmic and architectural co-design for CNN kernels running on vector architectures. The goal of the current study is to close this gap. Our goal is not to provide the fastest algorithm or to design a new vector architecture. Our goal is to study their combined impact to do a design space exploration of vector architecture and provide guidance for algorithm programmers, hardware designers, and compiler developers. 

This paper studies the interplay between algorithmic optimizations and micro-architectural parameter choices, demonstrating the trade-offs in co-designing CNNs and vector architectures. 
%To achieve this, first, we optimize the kernels found in convolutional layers on modern vector architectures and study the impact of these optimizations on performance. We then analyze the impact of micro-architectural parameters tuning on the optimized kernels.  
For our co-design study, we vectorize all the kernels of the convolutional layer from the Darknet framework~\cite{yolov3} on the RISC-V Vector and ARM-SVE architectures, using high-level intrinsics of the respective ISAs. We then optimize GEMM, the most time-consuming kernel, using various BLIS-like\cite{blis} techniques to reduce the pressure on the memory-subsystem, enforce contiguous memory accesses, and maximize the utilization of vector registers. We additionally optimize the Winograd algorithm from NNPACK \cite{nnpack}, with VLA vectorization on ARM-SVE, proposing a novel, inter-tile parallelism scheme.

We then use the gem5 \cite{binkert-gem5} simulator to assess the impact of tuning hardware parameters such vector lengths, vector lanes and L2 cache sizes, on the optimized kernels. We consequentially also assess the impact of integrating vector units tightly to the core, in the case of ARM-SVE, or as a decoupled vector architecture, in the case of RISC-V Vector.
%An exploration using the gem5~\cite{gem5benchmark,Gem5ARMSVE} simulator is conducted to assess the impact of our proposed combined study. 
Finally, we evaluate the performance of Winograd as an algorithmic replacement for im2col+GEMM. 
%The vectorized Winograd implementation is used to further used to implement the convolutional layer of the Darknet framework. 

In summary, we make the following contributions: 
\begin{enumerate}
\item %We study several optimizations including BLIS-like optimizations on ARM-SVE and RISC-V Vector architectures to achieve high performance for CNN inference network models. 
We demonstrate that not all algorithmic optimizations are beneficial to all different vector architectures, due to traits of their micro-architectural design. %For example, BLIS-like optimizations do not enhance the performance in the case of RISC-V Vector. 
To the best of our knowledge, this is the first work that shows the impact of different algorithmic optimizations with CNN kernels, on ARM-SVE and RISC-V Vector ISAs.

\item% We study the impact of architectural parameters with the optimized kernels on vector architectures. 
%Our co-design study shows that longer vector lengths and larger caches with low latency enhance the performance of vector architectures. Moreover, more vector lanes are beneficial, but only to longer vector lengths. 
%However,
We characterize the impact of hardware parameters on convolutional layers with im2col+GEMM, showing that longer vectors can improve the performance by up to 2.5$\times$, and larger caches %with low latency 
can further improve performance 
by up to 1.9$\times$ (i.e.~a total of almost 5$\times$), when compared to a 512-bit long vector architecture with 1MB of L2 cache. 
%We observe that larger caches have a higher impact on performance for long vectors compared to short vectors. Moreover, additional vector lanes can improve performance, but this is limited to long vectors.

%However, co-design with Winograd changes the observations with caches and shows the almost same impact with larger caches irrespective of the vector lengths
\item We present a novel, vectorized implementation of Winograd with ARM-SVE in a VLA manner, offering up to 1.35$\times$ and 1.5$\times$ higher performance to YOLOv3 and VGG16 network models respectively, compared to im2col+GEMM, on a single core of A64FX. To the best of our knowledge, this is the first implementation of Winograd utilizing the longer, 2048-bit vectors. Moreover, our co-design study on ARM-SVE shows that Winograd is less sensitive to the L2 cache size compared to im2col+GEMM. 
%We provide an optimized implementation of a convolutional layer  for the Winograd implementation from NNPACK in vector length agnostic (VLA) way, which can utilize the vector lengths up to 2048 bits on ARM-SVE. We are showing that Winograd performs 12\% better than im2col+GEMM for the VGG16 network model on A64FX processor. To the best of our knowledge, this is the first implementation of Winograd utilizing long vectors of 2048 bits. 

\end{enumerate}
The rest of this paper is organized as follows. Section~\ref{sec:background} offers background on vector architectures and CNNs. Section~\ref{sec:methodology} presents our experimental platforms and setup. Section~\ref{sec:optimizations} describe the algorithmic transformations and optimizations on the most-time consuming kernels of the convolutional layer, namely im2col+gemm and Winograd. Section~\ref{sec:hwtuning} details the hardware parameters we consider in our co-design study. Section~\ref{sec:evaluation} demonstrates our co-design study on the RISC-V and ARM-SVE architectures. 
%Section~\ref{sec:multicore} evaluates the YOLOv3 network model with our optimized kernels on the Fujitsu A64FX system. 
Section~\ref{sec:winograd} evaluates the Winograd kernel and Section~\ref{sec:conclusions} concludes the paper. 
% ====================
% # II. MMIC designs #
% ====================

\section{Background}
\label{sec:background}
\subsection{Vector Architectures}
\label{sec:vecarch}
Although long vector lengths were used in supercomputers in the past \cite{vectorarhitecturepast}, and short vectors later became popular in general-purpose architectures \cite{Intel, Arm-neon}, 
%Vector supercomputers with long vector lengths \cite{vectorarhitecturepast} were originally built to solve scientific problems. A new era of SIMD architectures began with short vectors, initially built for media streaming applications which later became popular in Digital Signal Processing (DSP) and general-purpose computing \cite{Intel, Arm-neon}. 
the high energy efficiency and scalable vector length of vector architectures has led to renewed interest in High-Performance Computing. While SIMD instruction set architectures with a fixed short vector length are available and commonly used for general purpose computing, introducing longer vector lengths requires a new ISA extensions, limiting portability. %instruction set Extension, limiting portability. %But each time there is interest in longer vector length, a new instruction set extension is required. % if there is a interest of longer vector length. 
To overcome this limitation, modern architectures such as RISC-V Vector~\cite{RISC-VV} and ARM-SVE~\cite{ARMSVE} offer vector length agnostic (VLA) ISAs that are portable across different hardware vector lengths. %Further details of these extensions can be accessed thourgh~\cite{RISC-VSpec, svespec}. 

\paragraph{RISC-V Vector Extension}
 This is the vector extension of the RISC-V Architecture, with 32 vector registers and a maximum supported vector length (MVL) of 16384 bits.
 %, i.e. a vector can hold 256 double precision or 512 single precision elements with a maximum element width (\textit{elen}) of 64 bits. 
Different vector lengths (\textit{vlen}) in powers of two, not exceeding the MVL (\textit{maximum vector length}), can be used. A vector instruction \texttt{vsetvl} determines the granted vector length (\textit{gvl}) at runtime, using the requested vector length (\textit{rvl}) in elements and the element width in bits (\textit{sew}) as input, handling VLA code generation with different \textit{vlen} is handled at runtime. %To vectorize loops with control flow, a vector register \texttt{v0} is used as an implicit operand in the vector instruction, to mask the instruction. 
The RISC-V Vector also supports strided-access, gather-load and scatter-store vector operations.  
%to support non-contiguous memory accesses.  %Further details can be accessed through RISC-V Vector spec~\cite{RISC-VSpec}

%, which is calculated using following expression:
%\begin{displaymath}
%    gvl = \min{(VLMAX, rvl)}, \textrm{where }VLMAX = vlen/sew \newline
%\end{displaymath}

%The returned \textit{gvl} does not exceed the maximum vector length. 

\paragraph{ARM Scalable Vector Extension (SVE)}  This is the vector extension of the ARMv8 architecture. The ARM-SVE ISA operates on 32 vector registers and 16 predicate registers. The supported MVL is 2048 bits, allowing to use different vector lengths at runtime, from 128-bit to 2048-bit in increments of 128-bits. Predicate registers are used for per-lane predication, where elements with active lanes get processed and inactive lanes either update the destination or leave the destination unchanged. For the scalar loop tail, ARM-SVE uses loop predication by masking out vector elements and by processing partial vectors.  ARM-SVE also provides gather-load and scatter-store vector instructions. % for non-contiguous memory accesses. % Further details can be accessed through~\cite{svespec}

\subsection{{Convolutional network models}}
\label{sec:networkmodel}
Convolutional neural networks are implemented in multiple deep learning frameworks. In this work, we focus on Darknet \cite{darknet13}, an open-source neural network framework written in C and CUDA. It supports many pre-trained convolutional network models for inference in various applications, such as object detection and image classification. These networks models consist of different types of layers, but the computationally dominant layer is the convolutional layer. In Darknet, a convolutional layer is built from the functions \texttt{GEMM}, \texttt{im2col}, \texttt{fill\_cpu}, \texttt{copy\_cpu}, \texttt{normalize\_cpu}, \texttt{add\_bias}, \texttt{scale\_bias} and \texttt{activate\_array}.   
%sonia Experiment
%\begin{itemize}
%\item \textit{GEMM}: Matrix multiplication, which dominates the execution %time
%\item \textit{Im2col}: Data transformation to implement convolution via %Matrix Mutiplication 
%\item \textit{Fill}: Fill array with values
%\item \textit{Copy}: Copy values of one array into another array
%\item \textit{Normalize}: Normalize all values of the array at filters %level
%\item \textit{Add bias}: Add biases to an array
%\item \textit{Scale bias}: Scale array values by multiplying them with biases.
%\item \textit{Activations}: Apply activation function to an array. 
%\end{itemize}

\paragraph{CNNs for object detection and image classification} 
A popular CNN for object detection is YOLOv3, which features 107 layers of five different types, out of which 75 layers are convolutional. A variant for the same task is YOLOv3-tiny, which features 23 layers, out of which 13 are convolutional. 

VGG16 %and Resnet50 
is an image classification CNNs. 
%where the dominant layer is the convolutional layer. 
VGG16 includes 25 layers, out of which 13 are convolutional and 3 are fully-connected layers. %Resnet50 includes 67 layers, out of which 50 are convolutional. 

\paragraph{Execution time breakdown for CNN inference} 
We profile the execution time of different kernels in the YOLOv3 network model, compiled with \texttt{clang} on the \textit{A64FX} system (see Section~\ref{sec:methodology} for details) and collect measurements using \texttt{perf}. Approximately 92\% of the total execution time is spent on computation for inference, while the remaining 8\% is used for setting up the network model. We exclude the time for setup, as it occurs only once, and calculate the percentage of time spent on each kernel with respect to the total computation time. The convolutional layer dominates execution, with \texttt{GEMM} consuming 93.4\% of the computation time.
%Table~\ref{tab:profiling} shows the profiling results for the YOLOv3 CNN inference network model. 

%We observe that the kernels of the convolutional layer consume most of the computation time, with \texttt{GEMM} consuming 93.4\% of the execution time. 

%\begin{table}
%\centering
%\caption{Profiling of YOLOv3 network model}
%\begin{tabular}{|c|c||c|c|} \hline
%\textbf{Functions}& \textbf{Profiling (\%)} & \textbf{Functions} & \textbf{Profiling (\%)}\\ \hline
%GEMM&93.4 & activate\_array&1.0\\\hline 
%other functions of convolutional layers&10\\\hline
%im2col&0.5& normalize\_cpu&2.5\\\hline
%fill\_cpu&0.12 & scale\_bias&0.1\\\hline
%copy\_cpu&0.12 & add\_bias&0.1\\\hline
%Maxpool Layer&1.8\\\hline
%\end{tabular}%
%\label{tab:profiling}
%\vspace{-1em}
%\end{table}

\paragraph{Convolutional layer implementations}
Our profiling results show that the convolutional
layer is the main building block of CNN network models. In Darknet, this layer is implemented using the im2col+GEMM algorithm, which is also the dominant kernel. We focus on the optimization of the generic im2col+GEMM algorithm, however, a convolutional layer can be implemented with multiple algorithms, as no "one-size-fits-all" convolution implementation exists \cite{cnnalgorithms}: \textit{Winograd} works best with convolutional layers with 3x3 or 5x5 kernel sizes\cite{ala2022winograd}, \textit{FFT} works best with layers with large kernel sizes, while the \textit{Direct} algorithm is better for 1x1 kernel sizes. 
%The im2col+GEMM implementation is generic and works with all types of convolutional layers, thus, in this work, we focus on its optimization for im2col+GEMM. 
%However, in CNN-based network models, most of the network models have convolutional layers with kernel sizes of 1x1, 3x3 or 5x5 ~\cite{Lu2017winograd}. 
We therefore additionally optimize the Winograd algorithm, using the NNPACK \cite{nnpack} package implementation, as in CNN-based network models, most of the network models have convolutional layers with kernel sizes of 1x1, 3x3 or 5x5 ~\cite{Lu2017winograd}.

\section{Methodology}
\label{sec:methodology}
\subsection{Hardware platforms}
Our experimental analysis focuses on RISC-V Vector and ARM-SVE architectures. For the exploration of hardware parameters, we simulate both architectures with gem5~\cite{binkert-gem5}, a cycle-accurate simulator that models the core pipeline, providing accurate timing predictions. For  ARM-SVE, we use the Fujitsu A64FX processor that implements the ARMv8-SVE architecture, to evaluate our algorithmic optimizations. 

\begin{table}[ht]
\centering
\caption{Hardware Platforms}
\begin{tabular}{|c|c|c|c|} \hline
 & \textbf{RISC-V Vector} & \textbf{ARM-SVE} & \textbf{A64FX}\\ 
 & \textbf{@ gem5} & \textbf{@ gem5} & \\ \hline
\textbf{ISA} & RISC-V Vector & ARM v8.2+sve & ARM v8.2+sve \\ \hline
%other functions of convolutional layers&10\\\hline
\textbf{Processor} & in-order & in-order & out-of-order \\\hline
\textbf{Clock Rate} & 2GHz & 2GHz & 2GHz\\\hline
\textbf{L1 Cache size} & 64kB, 4-way & 64kB,4-way & 64kB,4-way\\\hline
\textbf{L2 Cache size} & 1MB, 8-way & 1MB, 8way & 8-MB, 16-way\\\hline
%L2 cache latency & 12 & 12 & 37\\\hline
%simd-units & 8 & 4 & 4\\\hline
\textbf{Cache line size} & 64b & 64b & 256b\\\hline
\textbf{Prefetching} & No & No & Yes\\\hline
\textbf{Vector Length} & up to 16384-bit & up to 2048-bit & 512-bit\\\hline
\textbf{Vector Lanes} & up to 8 & proportional  & not configurable \\
& & to vector length & \\\hline
%Maxpool Layer&1.8\\\hline
\end{tabular}%
\label{tab:systems}
\end{table}

The specifics of the hardware platforms used for our experiments are described in Table~\ref{tab:systems}. We note that A64FX has 2 SIMD units, and the vector lengths are not reconfigurable, as this is an actual processor. We use a RISC-V fork of gem5~\cite{gem5benchmark} and the public version of the gem5 simulator~\cite{Gem5ARMSVE} with support for modeling vector architectures, for RISC-V Vector and ARM-SVE, respectively, in system call emulation (SE) mode. We configure gem5 with the in-order "MinorCPU" CPU model, with a frequency of 2GHz for the CPU and vector processor unit(VPU). The memory subsystem is configured with two levels of data cache. We note that, on RISC-V Vector, the VPU is connected to the L2 cache. A small VectorCache buffer of 2KB is used, through which the VPU reads and writes data from/to the L2 cache. However, on ARM-SVE, data for vector registers is accessed through the L1 cache itself. 

\subsection{Experimental setup}
We evaluate the YOLOv3 network models from the Darknet framework on a 768 $\times$ 576 pixels input image. To compile the models, we use LLVM \texttt{clang}~\cite{llvm-epi} cross-compiler v12.0.0 for RISC-V Vector,  LLVM \texttt{armclang} v20.3~\cite{ARMCLANG} for ARM-SVE @ A64FX and \textit{GCC} cross-compiler version 10.2 for ARM-SVE @ gem5. For both RISC-V Vector and ARM-SVE, we use the \texttt{-O3} optimization flag. To collect baseline results, we use the \texttt{-fno-vectorize} compiler flag in both compilers. Note that the baseline implementation of the network models in Darknet does not include any manual vectorization. 

To analyze the impact of the vector lengths, we vary the vector lengths in both simulated architectures from 512 bits up to 2048 bits on ARM-SVE and up to 16384 bits on RISC-V Vector, in powers of 2. To analyze the impact of on-chip parallellism on RISC-V Vector, we vary the number of vector lanes from 2 up to 8. To analyze the impact of cache parameters, we increase the L2 cache size on both simulated architectures from 1MB up to 256MB. We calculate the L2 cache latency using the latency of AMD Zen2 L2 (12 cycles @ 7nm tech) and extrapolating it to a cache size of 1MB, using the CACTI tool~\cite{CACTI}, resulting in a latency of 12 cycles. 

To collect time measurements, we perform 100 repetitions for all experiments on A64FX, achieving a 95\% confidence interval of the mean falling within 5\% of the average.
%For the extrapolation, we consider the following parameters: the technology node is 22 nm process, one read-write port is used per bank (8 banks total), and high performance (itrs-hp) transistors are selected. We use a constant multiplier to determine the latency for L2 cache size, which is calculated as 12 cycles for 1MB of L2 cache size.

%For baseline results, the respective architecture's clang compiler is used with "-O3" and "-fno-vectorize" compiler flags. Note that the baseline implementation of the network models in Darknet does not include any manual or auto-vectorization.  
%For experimental analysis of the performance we use the gem5 simulator~\cite{binkert-gem5}. We use gem5 for two reasons. First, gem5 is a cycle-accurate simulator that models the core pipeline in order to provide accurate timing predictions and analyse the impact of micro-architectural parameters on performance. Second, gem5 can be used to analyse both RISC-V and ARM-SVE which helps to perform the comparison of the vector ISAs. 
 %\subsection{Simulators and Compilers}
\label{sec:methodsim}

\section{Algorithmic Optimizations}
\label{sec:optimizations}
In this section, we focus on the algorithmic optimizations for im2col+GEMM for the convolutional layer. We additionally describe the optimization of the Winograd implementation of convolutional layers.

%For im2col with GEMM implementation we are optimizing the kernels of the convolutional layer of the naive Darknet. For Winograd implementation on convolutional kernel, NNPACK implementation is optimized on ARM-SVE and then used to implement Darknet's convolutional layer. 

\subsection{im2col+GEMM optimizations}
To maximize the attainable performance, we begin by vectorizing all kernels of the convolutional layer in Darknet with low-level intrinsic instructions of the respective ISAs on each of our experimental platforms. However, as discussed in Section~\ref{sec:background}, \texttt{GEMM} is the most time consuming kernel, and aside from vectorization, manual optimizations are necessary to extract the maximum parallelism out of im2col+GEMM. 

%Figure~\ref{fig:gemmpic} shows the im2col+GEMM implementation for the convolutional layer in Darknet.  
Assuming a convolutional layer with a $k \times k$ kernel size, on an input image of dimensions $k \times k \times c$, where $h$, $w$, $c$ are the height, width, and number of channels respectively, for $n$ number of filters,  GEMM takes as input, a weight matrix $M \times K$, and an input matrix $K \times N$, where $M =n$, $K = k \times k \times c$, and $N = h \times w$. 
%To simplify the notation, we hereafter denote the GEMM dimensions $M', N', K'$ as $M, N, K$, not to be confused with the dimensions of the image, the kernel size, and the number of filters. \\ 

%$M$ = number of filters (N), 
%$N$ = $H \times W$,
%and 
%K = $K \times K \times C$
%where $M$ equals to the number of $N$ filters, $N$, $K$ are %K Wight matrix works upon MxK matrix size and input matrix %works upon KxN matrix size where M, N and K are:
 
%M = Number of filters (N)

%N = Height * Width (H x W)

%K = Kernel size * kernel size * Input channel (K x K X C)

%%%sonia - experiemnt
%\begin{figure}[t]
%  \centering
%  \includegraphics[width=.9\linewidth]{im2col+gemm.png}
%  \caption{GEMM in convolutional layer with 3x3 kernel}
%  \Description{gemm in Yolov3}
%  \label{fig:gemmpic}
%\end{figure}

\begin{figure}[!h]
\small{
\begin{algorithmic}[1]
\STATE $i\gets 0$ , $j\gets 0$, $k \gets 0$
\FOR{$i\gets0$, $i< M$, $i++$ }
        \FOR{$k\gets0$, $k< K$, $k++$ }
                \STATE  A$_{alpha}$ = $alpha$ * A 
                \FOR{$j\gets0$, $j<N$, $j+=1$ }
                        \STATE C += A$_{alpha}$ * B
                \ENDFOR
        \ENDFOR
\ENDFOR
\end{algorithmic}
}
\caption{Naive implementation of GEMM}
\label{fig:naiveimp}
\vspace{-0.3em}
\end{figure}

Fig.~\ref{fig:naiveimp} shows the pseudocode for the naive implementation of GEMM, used in Darknet. Here, $A$ ($M \times K$) represents the weight matrix, $B$ ($K \times N$) represents the input matrix and $C$ ($M \times N$) represents the output matrix. 

To optimize GEMM, we follow two approaches. The first approach optimizes the \textit{3-loop implementation}, depicted in Fig.~\ref{fig:3loops}. The second approach tiles the matrices, resulting in a \textit{6-loop implementation} depicted in Fig.~\ref{fig:6loops}, where we apply optimizations. 
%\begin{enumerate}
%    \item Optimizations with 3-Loop implementation
%    \item Optimizations with 6-Loop implementation
%\end{enumerate}
%In this paper, we refer these optimization approaches as 3-Loops and 6-Loops implementations. 

We apply the following optimization to the \textit{3-loop implementation}:
i) vectorization with intrinsic instructions ii) contiguous memory loads/stores to/from vector registers, iii) loop reorder, and iv) loop unrolling. Loop reorder reduces the pressure on the memory subsystem, by maximizing the reuse of the vector registers. Loop unrolling hides the pipeline latency by maximizing the vector register utilization and increasing the parallelism in the algorithm. 

\begin{figure}[!h]
\small{
\begin{algorithmic}[1]%[Base algorithm with loop interchange used for vectorization]
\STATE $i\gets 0$ , $j\gets 0$, $k \gets 0$
\STATE long int gvl;
\FOR{$j\gets0$, $j<N$  }
        \STATE Calculate GVL using vsetvl vector intrinsic
        \FOR{$i\gets0$, $i< M$, $i+=unrollfactor$ }
                \STATE Load C Matrix in vector register
                \FOR{$k\gets0$, $k< K$, $k++$ }
                        \STATE Load B Matrix in vector register
                         \STATE  A$_{alpha}$ = $alpha$ * A[] \\
                         // if ALPHA=1 then skip multiplication\\
                         \STATE Broadcast A$_{alpha}$ in vector register 
                         \STATE vfmacc (VC,vaalpha, VB, gvl)
                \ENDFOR
                \STATE Store C Matrix from vector register VC
        \ENDFOR
        \STATE $j+=GVL$
\ENDFOR
\end{algorithmic}
}

\caption{Optimized 3-loop implementation of GEMM}
\vspace{-0.3em}
\label{fig:3loops}

\end{figure}
Figure~\ref{fig:3loops} shows the pseudocode for the optimized 3-loop implementation of the GEMM kernel. The loop in line 3 is incremented by the vector length $GVL$ to take advantage of VLA and the loop in line 5 is incremented by $unrollfactor$ to take advantage of loop unrolling, increasing parallelism and maximizing the register utilizations. Loops are reordered to reuse the loaded vector data as much as possible. Low level intrinsics are used to manually vectorize the algorithm. For the RISC-V Vector architecture, the vector length is calculated using the \texttt{vsetvl} intrinsic instruction. 
%and the vector length is calculated on the top \textit{j} loop counter.
%Inside the \textit{i} loop, matrix \textit{C} is loaded in vector registers according to the vector length. Then, in the inner-most loop $k$, matrix \textit{B} is loaded in the vector register. Next, matrix $A$ is scaled by the $alpha$ value and stored in $A_{alpha}$, then broadcasted to the vector register. If $alpha$ value is 1 then broacast the value of $A$ without scaling it by the $alpha$. 
Once matrices are loaded to the vector register, we use a fused multiply-add vector intrinsic \texttt{vfmacc} to calculate the multiplication and addition for the intermediate resultant matrix $VC$. 
\texttt{vaalpha}, a scalar value broadcasted to the vector register, is passed as the second parameter to the \texttt{vfmacc} intrinsic. The compiler internally uses vector-scalar multiply-add intrinsics and avoids the use of the broadcast intrinsic instruction. The resulting multiple multiply-add operations hide the pipeline latency.

Furthermore, we optimize the \textit{6-loop implementation}, where the original matrices in GEMM are tiled in blocks of dimensions $blockM$, $blockN$, $blockK$. We apply the following BLIS-like \cite{blis} optimizations: i) loop reorder, ii) matrix packing, iii) block size tuning, iv) loop unrolling, v) prefetching, and vi) vectorization using intrinsic instructions. 
We perform loop reorder and unrolling for the same reasons as in the \textit{3-loop implementation}. We pack matrices to facilitate contiguous memory accesses. We tune the block sizes to the size of the caches, in order to minimize memory accesses and maximize reuse. Finally, prefetching assists in hiding load latencies. 
%While optimizing the Figure~\ref{fig:vect} implementation,  we are using intrinsic instructions for vectorizing the inner-most loop and packing of matrices functions. 

\begin{figure}[!h]
\small{
\begin{algorithmic}[1]%[Base algorithm with loop interchange used for vectorization]
\STATE $i\gets 0$ , $j\gets 0$, $k \gets 0$
\STATE long int gvl;
\FOR{$j1\gets0$, $j1<N$, $j1+=blockN$ }
    \FOR{$k1\gets0$, $k1<K$, $k1+=blockK$ }
        \STATE Pack MatrixB
        \FOR{$i1\gets0$, $i1<M$, $i1+=blockM$ }
            \STATE Pack MatrixA
            \FOR{$j\gets0$, $j<blockN$,  }
                \STATE Calculate GVL using vsetvl vector intrinsic
                \FOR{$i\gets0$, $i< blockM$, $i+=unrollfactor$ }
                    \STATE Prefetch block of C matrix in L1 cache
                    \STATE Prefetch packed A matrix in L2 cache
                    \STATE Prefetch packed B matrix in L2 cache
                    \STATE Load C Matrix in vector register
                    \FOR{$k\gets0$, $k< blockK$, $k++$ }
                        \STATE Prefetch packed B matrix in L1 cache
                        \STATE Prefetch packed A matrix in L1 cache
                        \STATE Load packed B Matrix in vector register 
                        \STATE  A$_{alpha}$ = alpha * A[] //from packed A matrix\\
                                // if ALPHA=1 then skip multiplication \\
                        \STATE Broadcast A$_{alpha}$ in vector register 
                        \STATE vfmacc (VC,V$_{aalpha}$, VB, gvl)
                        ......
                    \ENDFOR
                    \STATE Store C Matrix from vector register VC
                \ENDFOR
                \STATE $j+=GVL$
            \ENDFOR
        \ENDFOR
    \ENDFOR
    \ENDFOR

\end{algorithmic}
}

\caption{Optimized 6-loop implementation of GEMM}
\vspace{-0.3em}
\label{fig:6loops}

\end{figure}
Fig.~\ref{fig:6loops} shows the pseudocode for the optimized 6-loop implementation of the GEMM kernel. The first three loops in  line numbers 3, 4, and 6 are incremented by block sizes $blockM$, $blockN$, and $blockK$, tuned to the architecture. Matrices are packed in lines 5 and 7, to facilitate contiguous cache access in the inner-most loop and facilitate prefetching. Matrix packing operations are also vectorized using the intrinsic instructions. 

The inner loops (lines 8 and 10) are incremented by $GVL$ (vector length) and $unrollfactor$, as in the 3-loop implementation, to make use of VLA and facilitate loop unrolling. Here, $VL \times unrollfactor$ is also called the macro-block size.  %where VL and 16. %are also termed as the macro-block sizes.
%Loop 4 in line 8 is incremented by VL to take the benefits of the vector length agnostic approach. Loop 5 in line 10 is incremented by 16 to facilitate the loop unrolling to increase the parallelization and maximize the register utilization.  
As in the 3-loop implementation, we perform loop reorder. Additionally, in this implementation, the blocks of matrix C are prefetched into the cache before storing them in the vector registers.  
%The inner-most loop in line 15 (inner-most loop) is incremented by 1. 
We also prefetch the $A$ and $B$ packed matrix data into the L1 cache. The remaining of this inner-most kernel is vectorized in the same way as in the 3-loop implementation. 
%using the intrinsic instructions. Load / Store intrinsic instructions with stride 1 is used to access the contiguous data from the cache. If alpha is 1 then we are skipping the extra floating point operation i.e., multiplication of alpha x A. For multiply-add operations, vector-scalar multiply-add intrinsic instructions are used. Multiple multiply-add operations will help in hiding the pipeline latency.

We note that the ability to prefetch depends on each platform. RISC-V Vector does not support prefetching, therefore any relevant intrinsic instructions are ignored by the compiler. In the case of ARM-SVE, the compiler generates the assembly instructions for prefetching, which take effect on the A64FX processor. However, due to the limited support for prefetching on gem5, in the simulated ARM-SVE on gem5, prefetch instructions are treated as no-ops.%\cite{gem5-prefetch}

\subsection{Winograd optimizations}
\begin{figure}[t]
  \centering
  \includegraphics[width=.8\linewidth]{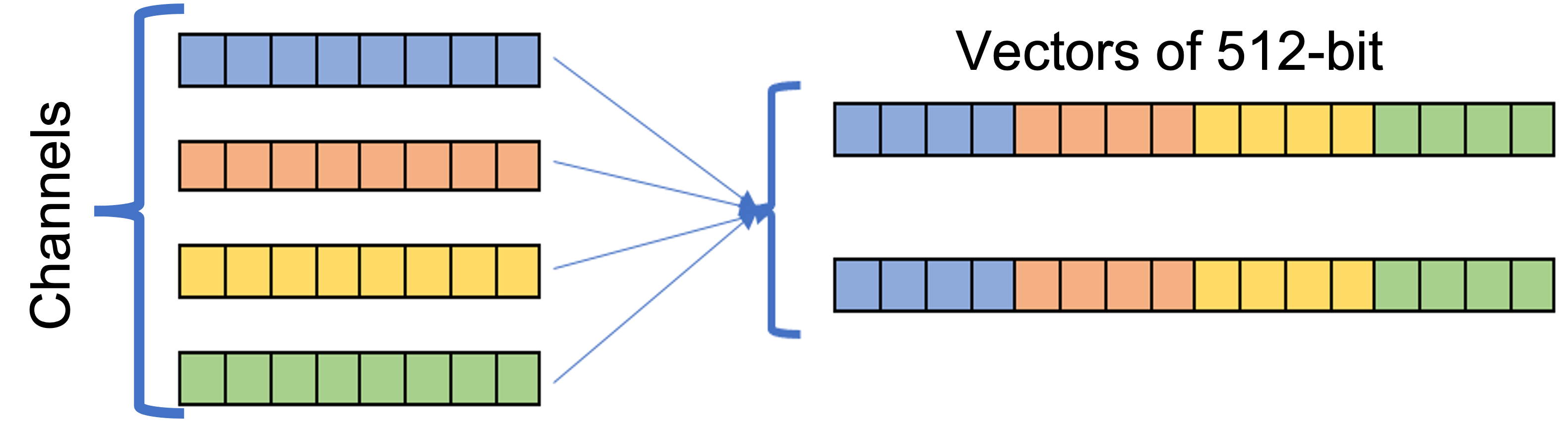}
  \caption{Inter-tile parallelism in Winograd}
%  \Description{gemm in Yolov3}
  \label{fig:winogradpic}
\vspace{-1em}
\end{figure}

As an alternative to im2col+GEMM, for convolutional layers with small filter sizes, we target the Winograd algorithm from the NNPACK \cite{nnpack} package. NNPACK and other implementations \cite{Maji2019EfficientWinograd, DongshengL12021Winograd} vectorize Winograd with ARM-NEON. However, we vectorize Winograd on top of the NEON implementation on NNPACK in VLA way, to utilize the longer vector lengths up to 2048-bit on ARM-SVE. The Winograd implementation %for a convolutional layer in NNPACK 
requires an input, weight, and output transformation and a tuple multiplication, and operates on a default tile size of 8$\times$8. Vectorizing the transformations with longer vector lengths would require a larger tile size, however, in this case, the numerical accuracy would drop. Therefore, we employ a scheme of inter-tile parallelism across the input/output channels by using an 8$\times$8 tile from each channel, which allows us to vectorize the transformation kernels using long vector lengths. Using 4 input/output channels with one row of 8$\times$8 tiles from each channel as shown in Fig.~\ref{fig:winogradpic}, we can utilize two 512-bit vector registers. To utilize longer vector lengths, we increase the number of 
input/output channels accordingly, e.g. 16 channels for 2048-bit vector registers.

The pseudocode in Fig.~\ref{fig:winograd} shows our inter-tile parallelization across the channels for the input transformation in Winograd. Lines 2 to 4 select the vector length, and determine the number of channels at runtime, in a VLA manner. For example, for a 512-bit vector length with 16 SP elements, the number of channels will be 4. If the number of channels is more than 4, inter-tile parallelism is enabled. Lines 6-16 create the buffers \texttt{buff1}, \texttt{buff2} to utilize the specified vector length. In Line 17, these buffers are used as a input for the SVE-vectorized input transformation kernel. We optimize the kernels for the weight and output transformation for longer vector lengths in a similar way, using the same inter-tile parallelization scheme, and applying the corresponding vectorized transformation (replacing the function in line 17).

\begin{figure}[!h]
\small{
\begin{algorithmic}[1]
\STATE $i\gets 0$ , $j\gets 0$, $k \gets 0$, $tileitr \gets 0$
\STATE $elements = 4$
\STATE $VL = svcntw()$. % get vector length
\STATE $interchannels$ = $VL / elements$ 
\IF{$count >= 4$} 
    \STATE $tiles = interchannels$
    \FOR{$tileitr\gets0$, $tileitr<count$, $tileitr +=tiles $ }
        \STATE //Below logic prepares the buffers for longer vectors
        \FOR{$k\gets0$, $k<tiles$, $k+=1$}
            \FOR{$i\gets0$, $i<8$, $i+=1$}
                \FOR{$j\gets0$, $j<4$, $j+=1$}
                    \STATE buff1[] = pack row-wise 0-4 elements of 8x8 tile
                    \STATE buff2[] = pack row-wise 4-8 elements of the same 8x8 tile
                \ENDFOR
            \ENDFOR
        \ENDFOR\\
        \STATE nnp\_iwt8x8\_3x3\_with\_offset\_sve\_vectorized \\
\begin{comment}

        \IF{$ROW ==8 and COULMN==8$}
        \STATE use the buff1 and buff2
        \ELSE put them in the block as per rows and coulmns and padd with zeros
        \ENDIF
        \STATE V00= svld1(buff[0])
        \STATE V01 = svld1(buff[1])
        \STATE ....
        \STATE ....
        \STATE V07 = svld1(buff[7])
        \STATE V10 = svld1(buff1[0])
        \STATE V11 = svld1(buff1[1])
        \STATE ...
        \STATE ...
        \STATE V17 = svld1(buff1[7])
        \STATE Compute wd0 := (V00 - V06) + 5.25 * (V04 - V02)
        \STATE compute wd1 := (V06 + V02) - 4.25 * V04
        \STATE compute wd2 := (V01 + V05) - 4.25 * V03
        \STATE compute wd3 := (V06 + 0.25 * V02) - 1.25 * V04
        \STATE compute wd4 := (V05 + 0.25 * V01) - 1.25 * V03
        \STATE compute wd5 := (V06 - 5.0 * V04) + 4.0 * V02
        \STATE compute wd6 := (V01 + 0.25 * V05) - 1.25 * V03
        \STATE compute wd7 := (V07 - V01) + 5.25 * (V03 - V05)
        \STATE d0 = wd0
        \STATE d1 =  wd1 + wd2
        \STATE d2 =  wd1 - wd2
        \STATE d3 =  wd3 + wd4 * const2
        \STATE d4 =  wd3 -  wd4 *  const2
        \STATE d5 = wd5 + wd6 * const2
        \STATE d6 = wd5 - wd6 *const2
        \STATE d7 = wd7
        \STATE Similarly d8-d15 are calculated with V10...V17
        \STATE transpose and merge four vectors d0, d1, d2, d3
        \STATE Same with d4 - d15
        \STATE Repeat step from line 31 to line 47 with the intermediate vectors to calculate final transposed values. 
\end{comment}
        \STATE Store the transposed data in their respective tiles across channels. 
    \ENDFOR
\ELSE
    \STATE $Tile = 1$
    \STATE nnp\_iwt8x8\_3x3\_with\_offset\_sve\_vectorized \\
\ENDIF

\end{algorithmic}
}
 \vspace{1em}
\caption{Input transformations code snippet from winograd showcasing the inter-tile parallelism across channels}

\label{fig:winograd}
\end{figure}

%To facilitates the longer vector lengths with these transformations, a big tile size is required. But winograd implementation increases the numerical accuracy with longer tile sizes. Therefore, all the transformation kernels are vectorized using inter-tile parallelism across the input/output channels by taking an 8x8 tile from each channel as shown in the Figure~\ref{fig:winogradpic}.
%Figure~\ref{fig:winogradpic} describes the inter-tile parallelism with 4 input/output channels by taking 8$\times$8 tiles from each channel and prepare two 512-bit vector registers. 
%We use two buffers/vector registers, following the Neon implementation logic in NNPACK, to perform the transformations on ARM-SVE. 
%Similarly our implementation uses 16 input/output channels to perform the inter-tile parallelism to prepare two 2048 bits of vector registers to utilize the longer vector length of 2048-bit. 

We additionally vectorize the tuple multiplication in a VLA way, which can use up to 2048-bits of vector length. To utilize the longer vector lengths for tuple multiplication, we increase the number of blocks for the GEMM kernel, using 16 blocks with 4 elements in each block. Therefore, there will be total 64 elements to utilize the maximum 2048-bit vector lengths. 
%\remove{The tuple multiplication is vectorized in the vector length agnostic way to utilize the different vector lengths such ad 512-bit or 1024-bit vector lengths. In this way, we offer an optimized Winograd implementation that can take advantage of VLA and utilize vector lengths up to 2048 bits. }

\section{Hardware Tuning}
\label{sec:hwtuning}
In this section, we detail our methodology to study the impact of tuning hardware parameters with the optimized kernels for CNN inference. We focus on three parameters: vector lengths, L2 cache sizes, and the number of vector lanes. To support CNNs, AI or scientific applications, latest chips are integrating longer vector lengths for fast processing. As the vector length increases, so does the pressure on the memory subsystem. Adding larger caches to alleviate the pressure can drastically increase the access time. Even if we assume constant access time, we still need to determine the exact cache size that is beneficial. Along with the cache sizes, there is a need to have more on-chip parallelism. However, bigger caches and more on-chip parallelism can influence the performance differently for different vector lengths. Therefore, it is important to study the trade-off between these micro-architectural parameters, as these hardware components occupy significant die area, while having an important influence on performance.  

%These three micro-architectural parameters are important in vector architectures, for two reasons. First, these hardware components occupy significant die area. Second, all three have an important influence in performance. 

%With the facts longer vector lengths can hide the latency and can benefit with the smaller cache sizes too and very long vector lengths occupy significant die space, it becomes important to know the trade-off between vector lengths and cache sizes. Another important consideration is that more the number of Vector lanes, more on-chip parallelism will be. More on-chip parallelism will provide more instruction level parallelism and will increase the start-up time. Therefore having more vector lanes can occupy die area and  the start-up overhead can influence in the performance differently with the different vector lengths. Therefore, it becomes important to know the tradeoffs between  

%This section covers studying the impact of tuning the hardware parameters with the optimized kernels. While doing the hardware parameters tuning, our study mainly focuses on vector lengths, L2 cache sizes, and vector lanes. These are the microarchitectural parameters that need studying because of two reasons. First, these hardware parameters take a significant amount of die space. Secondly, these parameters can impact the performance either positively or negatively. 

Vector length agnostic ISAs can work with different vector lengths without any modifications in ISA, hence making vector length a hardware parameters in designing vector architectures.
%Therefore, the vector length becomes a hardware implementation parameter in the design of vector architectures. 
With latest ISAs supporting very long vector lengths, this raises the question of \textit{how long the vector lengths should be}. Tuning the vector lengths to the demands of optimized CNN kernels can guide hardware designers in the selection of the appropriate vector lengths on future architectures. 

Longer vector lengths require more on-chip storage , which consequently may require larger cache to effectively handle locality.  Larger cache sizes can reduce the cache miss rate, 
%but they come with increased access times. Even if we optimistically assume that cache latency does not change, 
therefore the question raised is \textit{how large should the L2 cache be for different vector lengths}. This question also relates to the length of vector registers, since longer vector registers can lead to increased pressure on the memory subsystem.

%Longer vector lengths process more data per processor cycle, therefore require higher data rates from the main memory to the cache and registers. Larger cache sizes can reduce the cache miss rate, but they come with increased access times. Even if the access time, i.e. the cache latency, does not increase, the question raised is \textit{how large should the L2 cache be for different vector lengths}. This question also relates to the length of vector registers, since longer vector registers can lead to increased pressure on the memory subsystem. 
%Also, if vector registers are getting increased for the latest ISAs then again it impact the L2 cache size because of the increased pressure on memory subsystem. Therefore, it is important to know the impact of L2 cache size with the different vector length. 
 
The number of vector elements to be processed per cycle is determined by the available on-chip parallelism. To achieve this, additional pipelines can be added to a vector architecture. However, the question raised is \textit{how many vector lanes are required for different vector lengths}, as adding more pipelines increases the start-up overhead, which can potentially degrade the performance with short vector lengths.  
%Therefore, number of vector lanes is needed to study with different vector lengths. 

%Similarly other following parameters should be studied:
%\begin{enumerate}
%    \item How long the vector lengths should be?
%    \item how big the L2 cache last level cache should be with different %vector lengths?
%    \item How much on-chip parallelism with different vector lengths?
%\end{enumerate}
%Looking at all three parameters, there is a clear interrelation. The L2 cache size is related to the ability to load more data into the VPUs and process more data. More vector lanes are related in such as way that longer vector lengths can hide the pipeline latency and hide the start-up time. Therefore, it is important to tune L2 cache size and vector lanes in the function of vector lengths. 

To respond to the aforementioned questions, our analysis shows the trade-offs between these three micro-architectural parameters. 
%we consider a number of different vector lengths in powers of two, both for RISC-V Vector and ARM-SVE, that do not exceed the maximum vector length available. We additionally consider L2 cache sizes ranging from 1MB to 512MB in multiples of two. We keep the cache latency constant, assuming it is not impacted for cache sizes up to 512MB. Finally, we consider 2 to 8 vector lanes in multiples of 2. 
We highlight that other micro-architectural parameters, such as such as in-order vs. out-of-order cores, the core frequency, or the number of registers, can also be important, but are beyond the scope of this paper.

\section{Evaluation of im2col+GEMM}
\label{sec:evaluation}
In this section, we present the results of our co-design study of CNN inference on RISC-V Vector and ARM-SVE. We first showcase the impact of algorithmic optimizations and hardware parameter tuning on RISC-V Vector, using a single core. For ARM-SVE, we evaluate our algorithmic optimizations in detail on the A64FX processor, and perform the hardware parameter tuning on ARM-SVE @ gem5. %Finally, we present results on multiple cores on A64FX. 
In all experiments with gem5, we report performance in terms of execution cycles. We exclude cycles spent on the initialization phase, network setup, as this is a constant overhead, not incurred when continuously running inference over a stream of images. 

%and the hardware parameter tuning, with the optimized versions of the im2col This section shows the results of our co-design study.The co-design study has been done performed on first on RISC-V Vector and then on ARM-SVE. The co-design study has been done on the single core.  Performance in terms of execution cycles is reported for all the experiments with gem5. Note that the results do not include the time to setup the network, these cycles are subtracted from total cycles. The initialization phase of the application is a constant overhead which is not incurred when continuously running inference over a stream of images. The performance is evaluated on A64FX architecture compared to the ARM Performance Library(ARMPL). The study on A64FX includes the evaluation on the single core as well on the multi-cores.  

\subsection{Algorithmic optimizations with RISC-V Vector}
\label{AlgOptRISCV}
We first analyze the performance of the algorithmic optimizations for im2col+GEMM on RISC-V Vector @ gem5. To vectorize the inner-most kernels of the optimized 3-loop and 6-loop implementations, we use the EPI builtins ~\cite{EPI-builtin}. In the 3-loop implementation we have tuned the loop unrolling factor by utilizing up to 32 vector registers. Our study shows no significant improvement beyond utilizing 16 registers for RISC-V Vector. In fact, by utilizing the 32 register, we experienced a performance drop by $\sim$15\% due to register spilling. Therefore, we set the $unrollfactor$ as 16 for the 3-loop  and 6-loop implementations. Moreover, for the optimized 6-loop implementation, we tune the block sizes of the matrices, determined by the $blockM$, $blockN$, $blockK$ parameters, to fit the packed matrices into the L2 cache, as, on the RISC-V Vector architecture, the VPU is connected to the L2 cache. 

We simulate the first 4 convolutional layers of the YOLOv3 network on the RISC-V Vector @ gem5, with 1MB of L2 cache and 8 vector lanes, on a single core, using the optimized 3-loop implementation and the optimized 6-loop implementation, with different block sizes. The relative execution time of the 6-loop implementation over the 3-loop implementation is presented in Table~\ref{tab:yolov3opt}, for block sizes of different dimensions. We observe that the optimal block size for the 6-loop implementation is $16 \times 512 \times 128$, where the two implementations only differ by 2\%, a difference that is not significant in the simulated environment. 

Overall, the results indicate that the optimized 6-loop implementation does not offer any performance benefit over the optimized 3-loop implementation on RISC-V Vector, despite the BLIS-like optimizations. We attribute this to the following two reasons. First, the 6-loop implementation packs the matrices to facilitate contiguous cache accesses during the inner-most loop and prefetches the packed $B$ and $A$ matrices in the L2 and L1 caches. The rationale behind tuning the block size in BLIS-like optimizations is to fit matrix $B$ in the last-level cache (L2) and matrix $A$ in the L1 cache. However, in RISC-V Vector, the VPU is connected to the L2 cache. Therefore, data in the L1 cache is not directly accessed by the VPU and practically, the implementation benefits only from caching in L2. Additionally, as also explained in Section~\ref{sec:optimizations}, RISC-V Vector does not support prefetching, which is a desired feature in the 6-loop implementation, in order to hide the latencies associated with matrix packing. 

We conclude that BLIS-like optimizations do not boost the performance of convolutional layers on RISC-V Vector. 
%\remove{ We have additionally validated this conclusion by executing the first 20 layers of YOLOv3, using the optimal block size for the 6-loop implementation.} 
We highlight that, after vectorizing all the kernels of the convolutional layer and by optimizing the im2col+GEMM kernel with the 3-loop implementation, we observe 14 $\times$ higher performance compared to the naive baseline for the YOLOv3-Tiny network model.

\begin{table}[htbp]
\centering
\caption{Relative execution time of YOLOv3 (4 layers) with the optimized 6-loop implementation, compared to the optimized 3-loop implementation of im2col+GEMM on RISC-V Vector @ gem5}
\begin{tabular}{|c|c|} \hline
\textbf{Block sizes} & Normalized Performance \\
 \hline
128$\times$1024$\times$256 & 0.9\\
 \hline
16$\times$1024$\times$128 & 0.95\\
 \hline
16$\times$512$\times$128 & 0.98 \\
 \hline
16$\times$512$\times$256 & 0.96\\
 \hline
32$\times$512$\times$128 & 0.97 \\
 \hline
64$\times$1024$\times$128 & 0.95\\
 \hline
\end{tabular}
\label{tab:yolov3opt}
\end{table}

%\begin{figure}[ht!]
%  \centering
%  \includegraphics[width=.9\linewidth]{figs/TuningBlockSize.pdf}
%  \caption{Normalized performance of YOLOv3 (4 layers) with the optimized 6-loop implementation, compared to the optimized 3-loop implementation of im2col+GEMM on RISC-V Vector @ gem5.}
%  \label{fig:yolov3opt}
%  \vspace{-1em}
%\end{figure}

\subsection{Hardware parameters tuning with RISC-V Vector}
Using the optimized 3-loop implementation, which demonstrates the best performance on RISC-V Vector, we proceed our experimentation with tuning hardware parameters of the architecture. We experiment with the first 20 layers of the YOLOv3 model, out of which 15 are the convolutional layers. 
%This section covers the hardware micro-architecture tuning with the optimized kernels on RISC-V Vector architecture. In this section, we will be looking the impact on the performance with different vector lengths, different cache sizes and different vector lanes for YOLOv3 network models mainly with the first 20 Layers out of which 15 are the convolutional layers. 

\paragraph{Scalability with different vector lengths}

\begin{figure}
  \centering
  \includegraphics[width=\linewidth]{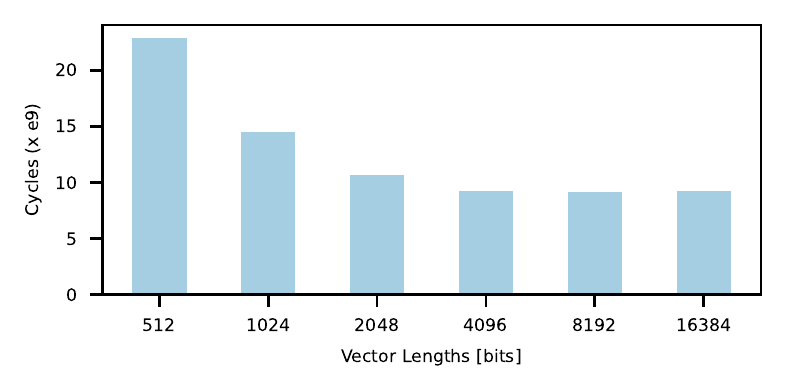}
  \caption{Impact of vector lengths on RISC-V Vector @ gem5 for YOLOv3 (20 layers), for constant L2 cache 1MB and 8 vector lanes.}
  \label{fig:yolov3onriscv}
   \vspace{-1em}
\end{figure}

Fig.~\ref{fig:yolov3onriscv} demonstrates the impact of different vector lengths on the performance of the convolutional layers on RISC-V Vector. For this experiment, we consider a fixed L2 cache size of 1MB and a fixed number of vector lanes, equal to 8 on gem5, varying only the vector length. We note that longer vector lengths hide the pipeline latency of vector lanes, thus any overheads associated to the start-up time become minimal. 
%However this study is carried out to highlight the impact of different vector lengths on the performance. 
Moving from 512-bit to 16384-bit vector lengths, the performance increases by 2.5$\times$, but effectively, the performance saturates beyond the 8192-bit vector length. To analyze this effect, we present the consumed average vector lengths and L2 cache miss rate, collected in gem5, in Table~\ref{tab:averageVL}. Although the 16384-bit vector length is almost fully utilized, the L2 cache miss rate increases significantly both for the 8192-bit and 16384-bit vector lengths. We therefore attribute this performance saturation to the increase in the L2 cache miss rate. Although longer vector lengths help in hiding latency, which should boost the performance without increasing the cache size, they also require more data to be processed per cycle, therefore to be transferred from the memory to the cache and then to the VPU, hence the increased L2 cache miss rate.
%This increase in the cache miss rate is a potential cause ese numbers indicates a big difference in cache miss rate. This is one of the potential reason which does not let IPC to be changed much from 8192-bit to 16384-bit.  

%\begin{figure}[ht!]
%  \centering
%  \includegraphics[width=.9\linewidth]{yolov3-tiny-risc-v-speedup.pdf}
%  \caption{Speedup of YoloV3-Tiny on RISC-V Vector by keeping L2 cache %size 1024kB}
%  \label{fig:yolov3tinyonriscv}
%\end{figure}

 \begin{table}[ht] 
\centering
\caption{Average vector length and L2 cache miss rate}
\begin{tabular}{|c|c|c|} 
\hline %inserting double-line 
\textbf{Vector length}                        &\textbf{YOLOv3}& \textbf{L2 cache miss rate(\%)} \\ \hline
512-bit      & 512&32              \\ \hline
1024-bit     & 1022.9&36          \\\hline
2048-bit   & 2041.9& 39\\\hline
4096-bit   & 4063.7 & 42\\\hline
8192-bit   & 8111.9 & 61\\\hline
16384-bit  & 15902.2 & 79\\\hline
\end{tabular}
\label{tab:averageVL}
\end{table}

\paragraph{Scalability with different L2 cache sizes}
We continue our hardware parameter tuning with the L2 cache sizes. We examine the impact of L2 caches for different vector lengths, since we have observed an increase in L2 cache miss rate for the 1MB cache as the vector length increases. For this experiment, we consider a fixed number of vector lanes equal to 8, on gem5. We expect a larger L2 cache to reduce the miss rates, however, one should note that larger caches come with increased access latencies and require more chip area.  %To observe the impact of L2 cache size on performance, the L2 cache size is increased from 1MB to 512MB. It is expected that larger cache size will reduce the L2 cache miss rate which results in performance improvement. But one should consider the research question \textit{how big the L2 cache size should be even if latency is not severely impacted as it may impact the die space area too}.
%Figure~\ref{fig:yolov3constantlongcacherisc} shows the scalability in performance in the function of cache sizes and vector lengths.
%\begin{enumerate}
%    \item How big the L2 cache size should be?
%    \item Impact of access time with bigger caches if latency is getting impacted
%\end{enumerate}
%  \begin{figure}[ht!]
%  \centering
%  \includegraphics{RISC-VV-Latency.pdf}
%  \caption{impact of very long caches with YOlOV3-Tiny on RISC-V Vector}
 % \Description{yolov3-tiny with long L2 cache size with constant Latency}
%  \label{fig:yolov3constantlongcacherisc}
%\end{figure}

%It is observed that performance speedup by 1.4$\times$ for smaller vector lengths and  1.5$\times$ with larger vector lengths. Another observation is that performance saturates at 256MB of cache size. 
 \begin{figure}[ht!]
  \centering
  \includegraphics[width=\linewidth]{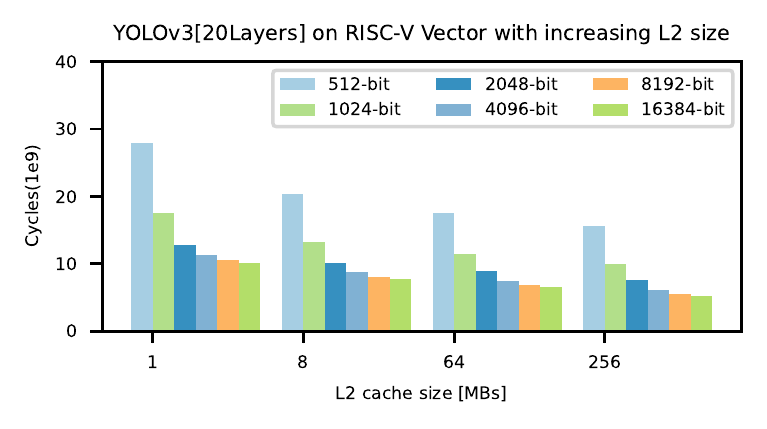}
  \caption{Impact of the L2 cache size on RISC-V Vector @ gem5 for YOLOv3 (20 layers), for 8 vector lanes.}
  
%  \Description{yolov3 with L2 cache size with constant Latency}
  \label{fig:yolov320longecachelongervl}
\vspace{-1em}
\end{figure}

Fig.~\ref{fig:yolov320longecachelongervl} shows the performance of YOLOv3 from 1MB to 256MB with different vector lengths. We observe that for vector lengths up to 4096 bits, the performance increases by 1.5$\times$ as we increase the L2 cache size. For the longer 8192-bit and 16384-bit vector length, the equivalent performance improvement is 1.7$\times$-1.9$\times$. We additionally observe that, with a 256MB L2 cache, performance improves by $\sim$5\% from 8192-bit vector length to 16384-bit vector lengths and L2 cache miss rates are 2.4\% and 2.6\% respectively. Therefore, we conclude that larger caches are beneficial, given that their latency remains low. Moreover, it is important to use larger L2 caches for longer vector lengths, but the performance gains of very long vector lengths are limited. 
Note that we have performed the same experiment on YOLOv3 using the optimized 6-loop implementation of im2col+GEMM, with block sizes tuned for an 8MB L2 cache, validating our conclusions regarding the L2 cache size tuning.

\paragraph{Scalability with different numbers of vector lanes}
We finalize our analysis of hardware parameters by tuning the number of vector lanes, i.e. the parallel SIMD units in the vector architecture that determine the on-chip parallelism. We examine the impact of this hardware parameter for different vector lengths, as increasing the number of vector lanes also increases the startup time; execution starts only after filling all the vector lanes. We note, however, that this analysis is limited by gem5 capabilities, which only allows to simulate up to 8 vector lanes. For this experiment, we consider a fixed 1MB L2 cache. Increasing the vector lanes from 2 to 8 with different vector lengths, we observe a performance improvement of $\sim$1.25$\times$ for the 8192-bit vector length. In the case of 512-bit, performance scales from 2 to 4 lanes, but becomes saturated beyond 4 lanes. We therefore conclude that additional vector lanes are more beneficial to longer vector lengths.

\subsection{Algorithmic optimizations with ARM-SVE}
\label{sec:arm-sve}
Similarly to RISC-V Vector, for ARM-SVE, we analyze the performance of the algorithmic optimizations for im2col+GEMM. For this, we use A64FX. We let the compiler to auto-vectorize all the kernels and manually vectorize kernels that the compiler fails to vectorize, such as normalization and activation. We manually vectorize the inner-most kernels of the optimized 3-loop and 6-loop implementations on SVE. 

Comparing the 6-loop implementation to the 3-loop implementation with ARM-SVE on A64FX, we observe a 2$\times$ performance improvement using the 6-loop, BLIS-like optimized GEMM kernel on the YOLOv3 network model. Unlike the case of RISC-V Vector, which poses the limitation of the VPU being attached to the L2 cache, on A64FX, the 6-loop implementation is able to take advantage of the caches and improve the performance of GEMM. Moreover, since prefetching is a hardware feature of A64FX, the prefetching instructions boost the performance of the 6-loop implementation. We note, however, that the 6-loop implementation outperforms the 3-loop implementation by 15\% on ARM-SVE @ gem5 which does not support prefetching, with a 512-bit vector length.
%e additionally note that the 6-loop implementation could potentially be performing better due to the existence of a prefetcher on A64FX, however, we have validated that  although this could also 
Comparing the optimized 6-loop implementation to the naive GEMM in Darknet, we observe a $\sim$32$\times$ performance improvement for YOLOv3 on A64FX.

\paragraph{Per-layer sustained performance}
We assess the sustained performance of the convolutional layers in YOLOv3, with respect to their arithmetic intensity, as per the roofline model, on A64FX, using our optimized kernels. YOLOv3 has 75 convolutional layers, but some layers work on the same input sizes. We therefore consider the 14 discrete convolutional layers which work with discrete matrix sizes, and compute the arithmetic intensity (AI) per layer as follows: 
\begin{eqnarray*}
{AI} & =\frac{Arithmetic Operations}{Bytes}
      =\frac{2\times M\times N\times K}{4\times( M\times N+ K\times N+M\times K)}
\end{eqnarray*}
where $M$, $N$, $K$ correspond to the sizes of the weight, input and output matrices. We showcase the results in Table~\ref{tab:AI}. We note that the peak performance of a single A64FX core is 62.5 GFLOPs. The results indicate that some layers have low AI and sustained performance, especially the layers with small $M$ and $K$ values, i.e., small weight matrix size. There is additional room for performance improvement for these layers, which is, however, beyond the scope of this paper, where we optimize kernels focusing primarily on portability across ISAs with VLA vector extensions. %However, the purpose of our algorithmic optimizations is not to offer the fastest kernels but to have highly optimized, portable kernels that leverage the long vector architectures and can facilitate our co-design study. 

%out which some of the layers work upon same type of data. For our per-layer analysis, we have taken the layers which work upon size of data.  
%The analysis with the correlation of Arithmetic Intensity(AI) and sustained performance as per the roofline analysis is studied per convolutional layer basis to analyze the achieved performance after applying all the optimizations. All the convolutional layers of CNN work upon the different sizes of matrices during the GEMM kernel. Therefore, we have calculated AI in table ~\ref{tab:AI} and achieved GFlops in figure~\ref{PEAKvssustained} of each convolutional layer to analyze per-layer performance with our implementation on a64FX.

%\begin{figure}[ht!]
%   \includegraphics[width=8cm]{peakvssustaied-yolov3}
%    \caption{Per-layer Performance for each layer}
%    \label{PEAKvssustained}
%\end{figure} 

 \begin{table}[t!] 
\centering
\caption{Arithmetic Intensity and Sustained performance of YOLOv3 convolutional layers on A64FX}
\begin{tabular}{|c|c|c|c|c|c|} 
\hline %inserting double-line 

\textbf{Layers}      & \textbf{M}&\textbf{N}&\textbf{K}&\textbf{AI}&\textbf{\% of Peak}              \\ \hline
L1    & 32& 369664&27&7.32&46           \\\hline
L2   & 64& 92416&288&26&72\\\hline
L3   & 32 & 92416&64&11&50\\\hline
L5   & 128 & 23104&576&52&77\\\hline
L6  & 64 & 23104&128&21&70\\\hline
L10  & 256 & 5776&1152&101&81\\\hline
L11 & 128 & 5776&256&42&75\\\hline
L38  & 256 & 1444&512&76&82\\\hline
L44  & 1024 & 361&4608&126&83\\\hline
L45  & 512 & 361&1024&88&78\\\hline
L59  & 255 & 361&1024&65&75\\\hline
L61  & 256 & 1444&768&85&91\\\hline
L62  & 512 & 1444&2304&162&83\\\hline
L75  & 255 & 5776&256&63&75\\\hline
\end{tabular}
\label{tab:AI}

\end{table}

%YOLOv3 has 75 convolutional layers out which some of the layers work upon same type of data. For our per-layer analysis, we have taken the layers which work upon size of data. Table~\ref{tab:AI} shows that every convolutional layer has different Arithmetic Intensity(AI). In the table M, N, K represent the matrix sizes of wight, input and output matrices as discussed in figure~\ref{fig:gemmpic} in the section~\ref{sec:optimizations}. The arithmetic intensity of each layer is calculated using following formula:
%\begin{eqnarray}
%{AI} & =\frac{Total Arithmetic %operations}{total bytes}\\
 %    & =\frac{2\times M\times N\times K}{4\times( M\times N+ K\times N+M\times K)}
%\end{eqnarray}

%This AI study also highlights that some layers have low AI and sustained performance. Especially the layers with small M and K values, i.e., small weight matrix size. We conclude that overall there is room for some improvement for the layers. However, our purpose is not to be the fastest but to have the high optimized kernels so that the hardware parameters tuning study can be performed on the top of high optimized kernels.

\subsection{Hardware parameters tuning with ARM-SVE}
\begin{figure}[!t]
  \centering
  \includegraphics[width=\linewidth]{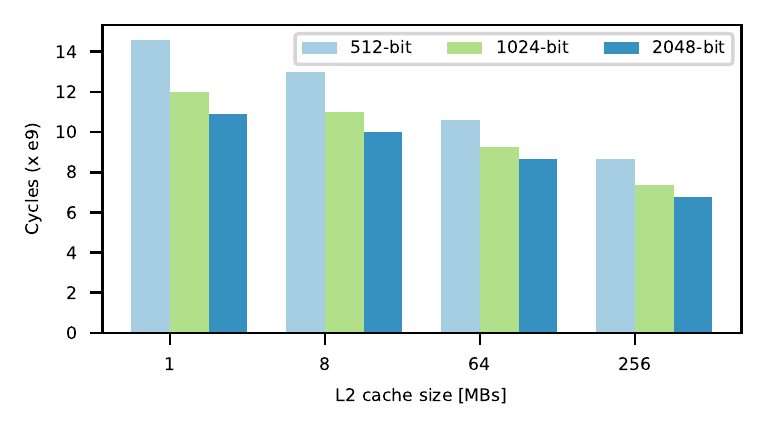}
  \caption{
  Impact of vector lengths and L2 cache size on ARM-SVE @ gem5 for YOLOv3 (20 layers).}
%  \Description{yolov3 with L2 cache size}
  \label{fig:yolov320llongercachearmsve}
  \vspace{-1.3em}
\end{figure}

Similarly to RISC-V Vector, we study the impact of micro-architectural parameters with ARM-SVE on gem5. 
%\sonia{We note that ARM-SVE did not benefit after adding more lanes, we focus on tuning the vector length and the L2 cache size. }  
As ARM-SVE on Gem5 sets the number of vector lanes proportional to the vector length, we focus only on tuning the vector length and the L2 cache size. We do not further tune the block sizes of the 6-loop implementation, as they fit in the smallest simulated cache. Fig.~\ref{fig:yolov320llongercachearmsve} shows the impact of different vector lengths and L2 cache sizes on the performance of the first 20 layers of YOLOv3. We observe that, for a cache of 1MB, moving from 512-bit to 2048-bit vector lengths, the performance improves by 1.34$\times$. Additionally, similarly to RISC-V Vector, performance benefits from larger caches, with a performance improvement of 1.6$\times$ as we increase the L2 cache size from 1MB to 256MB for 2048-bit vector length. Our findings for ARM-SVE agree with our observations for RISC-V Vector. Our optimized kernels can benefit from longer vectors and larger cache sizes, which can significantly boost the performance of CNN inference on vector architectures.   

%sthat our observation the performance on A64FX also  , we proceed to tune L2 cache size in the function of vector lengths. The performance analysis has been done with the 3-loop and 6-loop implementation on ARM-SVE @ gem5 too. we have kept the block sizes same as of block sizes used for A64FX. A performance improvement of $\sim$15\% of performance improvement with 6-Loop implementation compared to 3-Loop implementation is observed with 512-bit vector length.  Further, the 6-Loop implementation is used for the tuning of the L2 cache sizes in the function of different vector lengths. For tuning of the L2 cache sizes, we are not changing the block sizes. This is because multiple blocks can reside in the larger caches and will help in scaling the performance with the larger cache sizes. Figure~\ref{fig:yolov320llongercachearmsve} shows the impact on performance in the function of vector lengths and L2 cache size. Figure~\ref{fig:yolov320llongercachearmsve} shows the performance speedup of 1.34$\times$  from 512-bit to 2048-bit vector lengths with 1MB and performance improvement of 1.6$\times$ with longer caches from 1MB to 256MB.  

\section{Evaluation of Winograd}
\label{sec:winograd}
As explained in Section~\ref{sec:optimizations}, we vectorize the transformation and tuple multiplication kernels of Winograd in a VLA way, using intrinsic instructions on ARM-SVE. Our kernels adapts the different vector lengths and can be executed with 512-bit, 1024-bit and 2048-bit vector lengths. We use these kernels in Darknet, to implement convolutional layers with kernel sizes of 3$\times$3 and stride 1 and 2. For convolutional layers of different kernel sizes, we fall back to our optimized im2col+GEMM. 

For our Winograd implementation on ARM-SVE, we use intrinsics to create tuples of four vectors and then transpose these vectors. On RISC-V Vector, currently, no specific intrinsics are available to perform these operations.  %Some builtins for creating tuples of two vectors have been added in the latest EPI-Builtins, which target the 1.0 version of the vector extension. However, Gem5 is based on the 0.7 version and will be incompatible to these latest builtins.
We therefore implemented a solution that uses temporary buffers and additional store and gather-load intrinsics. This however limits the performance improvement and the potential insights of running Winograd on the RISC-V Vector extension with very long vectors. Because of this reason, we do not include RISC-V results in the Winograd analysis.
% This works, but could we say something about how we believe the results will look like? 

%We evaluate Darknet with the vectorized Winograd on ARMIE emulator with 512-bit, 1024-bit and 2048-bit vector lengths. 
\subsection{Algorithmic optimizations with ARM-SVE}
We evaluate the performance of the optimized Winograd implementation in Darknet on the A64FX processor. As a baseline for comparison, we use our optimized im2col+GEMM. We note that a naive implementation of Winograd is slower than using the naive implementation of im2col+GEMM, therefore we use our optimized im2col+GEMM as the baseline for comparison. A primary analysis revealed that the weight transformation is a major bottleneck, but it can be performed offline for inference. After excluding the weight transformation time, we achieve a speedup of 1.5$\times$
compared to im2col+GEMM for VGG16, where all convolutional layers use 3$\times$3 kernel-sized filters. For YOLOv3, where 38 out of the 75 use 3$\times$3 kernel-sized filters, the equivalent speedup is 1.35$\times$. Out of these 38 layers, the 32 with stride 1 perform 2.4$\times$ better with Winograd compared to im2col+GEMM, while for the 6 layers with stride 2, Winograd is 1.4$\times$ slower than im2col+GEMM. The remaining layers use 1$\times$1 kernel-size filters and default to im2col+GEMM. 
%Further investigation reveal that for YOLOv3 model, non-strided layers with Winograd performs 2.4$\times$ better than im2col+GEMM.Only 6 layers out of 75 convolutional layers with stride 2 are 1.4$\times$ slower with Winograd compared to im2col+GEMM.
We therefore conclude that our optimized Winograd algorithm offers significant performance improvement for layers with stride 1, however, different algorithmic optimizations are required to achieve high performance for layers with stride 2. Still, convolutional layers require careful algorithmic selection related to the kernel sizes and strides. 
%With these findings, we conclude that algorithmic selection should be made as per convolutional layer parameters such as stride and kernel sizes else more efforts are required to further optimize the Winograd for strided convolutional layers.  }%Note that we do not compare our implementation against the ARM-NEON implementation provided in NNPACK and other libraries, which only use 128-bit vector lengths, significantly shorter than the vector lengths available on ARM-SVE. 

%The reason is that the comparison will not be valid between 128-bit vector length and 512-bit or more
\subsection{Hardware parameter tuning with ARM-SVE}
Similarly to our approach for im2col+GEMM, we study the impact of hardware parameters on the performance of our optimized Winograd algorithm for ARM-SVE, using Gem5. As indicated by our evaluation on A64FX, we use Winograd for all convolutional layers with 3$\times$3 kernel sizes and stride 1, and default to our optimized im2col+GEMM implementation for all other cases. In particular, we study the impact of the L2 cache size, ranging from 1MB up to 256MB, and the impact of different vector lengths, i.e. 512-bit, 1024-bit and 2048-bit. The number of vector lanes is propotional to vector lengths.  
%We used the same A64fx Winograd binaries to study the hardware parameters impact on performance using gem5 for ARM-SVE. 
 
%For these experiments, we consider vector lengths as 512-bit, 1024-bit, and 2048-bit in the function of L2 cache sizes from 1MB to 256MB on Gem5.

\begin{figure}[!t]
  \centering
  \includegraphics[width=\linewidth]{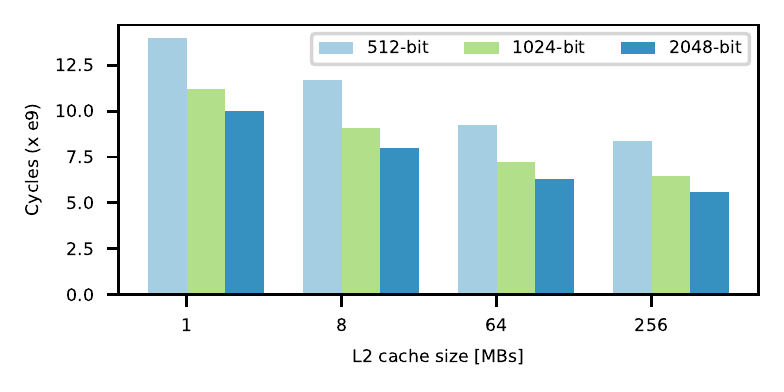}
  \caption{
  Impact of vector lengths and L2 cache size with Winograd on ARM-SVE @ gem5 for YOLOv3 (20 layers).}
%  \Description{yolov3 with L2 cache size}
  \label{fig:yolov320llongercachearmsvewinograd}
  \vspace{-1.3em}
\end{figure}

\begin{figure}[!t]
  \centering
  \includegraphics[width=\linewidth]{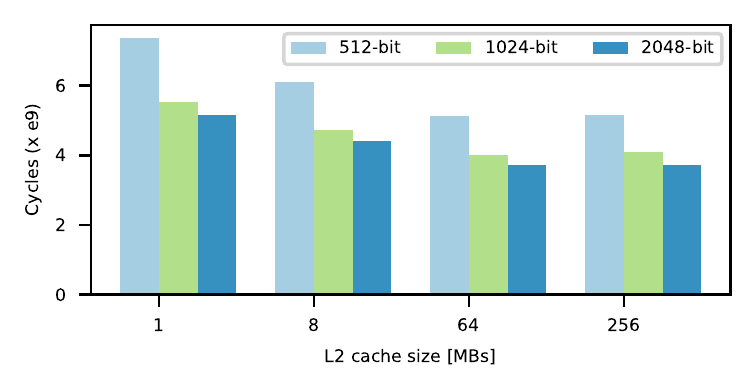}
  \caption{
  Impact of vector lengths and L2 cache size with Winograd on ARM-SVE @ gem5 for VGG16.}
%  \Description{yolov3 with L2 cache size}
  \label{fig:vgg16longercachearmsvewinograd}
  \vspace{-1.3em}
\end{figure}

We showcase the results of our analysis for the first 20 layers of YOLOv3 in Fig.~\ref{fig:yolov320llongercachearmsvewinograd}, and for VGG16 in Fig.~\ref{fig:vgg16longercachearmsvewinograd}. For both network models, for an L2 cache of 1MB, we observe a performance improvement of 1.4$\times$ as we increase the vector lengths from 512 to 2048 bits, due to increased throughput and decreased pressure on the memory subsystem. 

Evaluating the impact of L2 cache sizes, we observe that, for the first 20 layers of YOLOv3, performance improves by 1.75$\times$ for all vector lengths, when increasing the caches from 1MB to 256MB. For VGG16, the performance improves by ~1.4$\times$ from 1MB to 64MB, but the network does not benefit from a larger cache. We note that all layers in VGG16 use Winograd, which has smaller cache requirements compared to im2col+GEMM, whereas several YOLOv3 layers invoke im2col+GEMM. As a conclusion, longer vectors are highly beneficial to the performance of Winograd-enabled convolutional layers and networks. With respect to the L2 cache size, our optimized Winograd algorithm does not have high cache requirements, and therefore is able to perform well with moderately large L2 cache sizes.

We finally compare the performance of VGG16 using Winograd, compared to im2col+GEMM, with different vector lengths of 512, 1024 and 2048  bits, with 1MB of L2 cache. The performance improves by 1.4$\times$, 1.5$\times$, and 1.3$\times$ respectively, compared to im2col+GEMM, for the different vector lengths, showing that Winograd is a good alternative to im2col+GEMM for any vector length.

% using Winograd-enabled layers, compared to im2col+GEMM,  At first both network models are executed with 512-bit, 1024-bit and 2048-bit with 1MB of L2 cache size and compared the performance achieved with im2col+gemm. For YOLOv3[20 Layers], we observe a performance improvement of 4\% - 9\% for 512-bit, 1024-bit, and 2048-bit using Winograd. We digged out the reason that Out of 15 convolutional layers, 3 layers are with stride 2, 6 layers are with 1x1 kernel size, the very first layer has the number of channels as 3 and only the remaining 5 layers contribute to improving the performance. For VGG16, where all the layers works with 3x3 kernel size and stride as 1 , we observe a performance improvement of 1.5$\times$, 1.3$\times$, 1.3$\times$ with 512-bit, 1024-bit, and 2048-bit respectively.  %YOLOv3 with first 20 layers and VGG16 full network model  are used for the simulation purpose with different VLs with L2 cache upto 256MB. 
%Further we observe the scalability in the function of vector lengths and cache sizes.  

%\section{Related Work}
%\label{relatedwork}
%\input{relatedwork}
%\section{Programmability comparisons of ISAs}
%\label{sec:programmability}
%\input{programmability}

%\section{AREA and Power estimation}
%\label{areapower}
%\input{areapower}

\section{Conclusion}
\label{sec:conclusions}
In this paper, we presented a hardware and software co-design study of CNN inference on modern vector architectures with variable vector lengths. Focusing on the most time-consuming kernels in convolutional layers, we have developed efficient, VLA-vectorized, optimized implementations of im2col+GEMM and the Winograd algorithm. 

Experimenting with two different ISAs, RISC-V Vector and ARM-SVE, we conclude that certain optimizations are not portable across vector architectures, and highlight the following portable optimizations: i) maximize utilization/reuse of vector registers, ii) use unstrided load/store instructions, for contiguous memory accesses, iii) use multiple multiply-add instructions to hide the pipeline latency. We additionally conclude that longer vector lengths improve performance even with smaller caches, however larger caches with low latencies can help minimize any adverse effects from increased cache misses. Finally, more vector lanes can hide the pipeline and startup latency for longer vector lengths. 

Our algorithmic optimizations using VLA ISAs for im2col+GEMM improve the performance of CNN inference by 14$\times$ for YOLOv3-Tiny on RISC-V Vector and by 32$\times$ for YOLOv3 on ARM-SVE, compared to the naive implementation of im2col+GEMM in Darknet. Our vectorized Winograd algorithm offers additional performance improvement of 1.35$\times$ and 1.5$\times$ to YOLOv3 and VGG16 respectively, while having lower cache requirements. 

We believe that our work is useful to programmers, hardware designers and compiler developers. In the future, we aim to extend our algorithmic optimizations for vector architectures to more kernels in DNN inference and examine additional, influential architectural and micro-architectural features.

% ==========================
% # III. Measurement Setup #
% ==========================

% ================== Circuit B ==================

% ==================
% # Conclusion #
% ==================

% ==================
% # Acknowledgment #
% ==================
% use section* for acknowledgment
%\section*{Acknowledgment}
%For the Summary paper submission only, no %acknowledgements are allowed. 

% ==============
% # REFERENCES #
% ==============
\bibliographystyle{IEEEtran}
\bibliography{IEEEabrv,biblio_rectifier}

\end{document}